\documentclass[a4paper,reqno]{amsart}
\usepackage[text={5.2in,8in},centering]{geometry}
\usepackage[backrefs]{amsrefs}
\usepackage{amssymb}
\usepackage{enumerate}
\usepackage{graphicx}
\usepackage{graphics}
\usepackage{mathrsfs}

\DeclareMathAlphabet{\mathpzc}{OT1}{pzc}{m}{it}

\usepackage{color}
\definecolor{trustcolor}{rgb}{0.71,0.14,0.07}

\numberwithin{equation}{section}
\theoremstyle{plain}
\newtheorem{theorem}{Theorem}[section]

\newtheorem{cor}[theorem]{Corollary}
\newtheorem{lemma}[theorem]{Lemma}
\theoremstyle{remark}

\newtheorem*{quest*}{Question}
\newtheorem*{remark*}{Remark}
\theoremstyle{definition}
\newtheorem{definition}[theorem]{Definition}
\newtheorem*{definition*}{Definition}
\newtheorem*{notation*}{Notation}
\newtheorem*{notations*}{Notations}
\providecommand{\B}{\mathbf}

\providecommand{\C}{\mathcal}
\providecommand{\CS}{\mathscr}

\providecommand{\D}{\mathbb}

\newcommand{\ee}{\mathrm{e}}

\providecommand{\esm}[1]{\D{E}\left[ #1 \right]}

\providecommand{\prob}[1]{\D{P}\left\{ #1 \right\}}

\DeclareMathOperator{\dist}{dist}
\DeclareMathOperator{\diam}{diam}

\DeclareMathOperator{\card}{card}

\DeclareMathOperator{\one}{\mathbf{1}}

\def\half{{\frac{1}{2}}}

\def\myset#1{{\left\{\,#1\,\right\}}}
\def\pt{{\partial}}

\def\Lam{{\Lambda}}

\def\BLam{\B{\Lambda}}
\def\om{{\omega}}
\def\eps{{\epsilon}}
\def\gam{{\gamma}}

\def\DR{\D{R}}
\def\DN{\D{N}}
\def\DZ{\D{Z}}
\def\DC{\D{C}}
\def\DD{\D{D}}

\def\rS{{\rm S}}

\def\uB{\B{B}}
\def\uG{\B{G}}

\def\uS{\B{S}}

\def\cJ{\C{J}}
\def\cM{\C{M}}

\def\cR{\C{R}}
\def\cS{\C{S}}
\def\cSt{\widetilde{\C{S}}}

\def\csX{\CS{X}}
\def\cW{\C{W}}

\def\uu{\B{u}}
\def\uv{\B{v}}

\def\ux{\B{x}}

\def\ffi{{\varphi}}

\def\lam{{\lambda}}

\def\tI{{\tilde{I}}}
\def\tF{{\tilde{F}}}

\def\qt{{\widetilde{q}}}

\def\s#1{{\bf (S.{\mbox{\boldmath${{#1}}$}})}}
\def\ss#1{{\bf (SS.{\mbox{\boldmath${{#1}}$}})}}
\def\ms#1{{\bf (MS.{\mbox{\boldmath${{#1}}$}})}}

\def\sssk#1{{\bf (SS.{\mbox{\boldmath${{#1}}$}})}}
\def\mssk#1{{\bf (MS.{\mbox{\boldmath${{#1}}$}})}}

\def\Wone{{\bf (W)}}

\def\diy{\displaystyle}

\def\ketbra#1#2{{ | {#1} \rangle \langle {#2}| }}

\def\mymax#1{{ \truc{\max} {} {#1}}}

\def\truc#1#2#3{\smash{\mathop{\,\, #1 \,\, }\limits^{#2}_{#3}}}

\def\tto#1{\smash{\mathop{\,\,\,\, \longrightarrow \,\,\,\, }\limits_{#1}}}
\def\funmapto#1#2#3#4#5{#1:\, & #2 &\rightarrow  &#3 \\&  #4 &\mapsto  &#5}

\def\be{\begin{equation}}
\def\ee{\end{equation}}
\def\ba{\begin{array}{l}}
\def\ea{\end{array}}

\def\bal{\begin{aligned}}
\def\eal{\end{aligned}}

\begin{document}
\title[Localization with Less Larmes: Simply MSA]
{Localization with Less Larmes: Simply MSA}
\author[V. Chulaevsky]{Victor Chulaevsky$^1$}
\address{$^1$D\'{e}partement de Math\'{e}matiques\\
Universit\'{e} de Reims, Moulin de la Housse, B.P. 1039,\\
51687 Reims Cedex 2, France\\
E-mail: victor.tchoulaevski@univ-reims.fr}
\date{\today}
\begin{abstract}
We give a short summary of the fixed-energy Multi-Scale Analysis (MSA) of the Anderson tight binding model in dimension $d\ge 1$ and show that this technique admits a straightforward extension to multi-particle systems. We hope that this short note may serve as an elementary introduction to the MSA.
\end{abstract}

\definecolor{db}{rgb}{0.0,0.0,0.50}
\definecolor{db1}{rgb}{0.0,0.0,0.90}

\maketitle
\section{Introduction. } \label{sec:intro}

In this paper\footnote{The present text is an improved version of an earlier manuscript (version 1) written in 2008. Specifically, we give here a sharper decay bound for $(\ell,q)$-subharmonic functions; cf. Lemma \ref{lem:SubH.1}.}, we study spectral properties of random lattice Schr\"{o}dinger operators at a fixed,
but arbitrary, energy $E\in\DR$, in the framework of the MSA.
The idea of the fixed-energy scale induction goes back to
\cite{FS83}, \cite{Dr87}
\footnote{I thank A. Klein and T. Spencer for pointing out the PhD thesis by H. von Dreifus
\cite{Dr87} (scientific advisor T. Spencer) where the fixed-energy approach had been used.}
and \cite{S88}.
While fixed-energy analysis alone does not allow to prove spectral
localization, it provides a valuable information. Besides, from the physical point of view, a sufficiently rapid decay, with probability one, of Green functions in finite volumes, combined with the celebrated Kubo formula  for the zero-frequency conductivity $\sigma(E)$, shows that $\sigma(E)=0$ for the disordered systems in question.

The main motivation for studying in this paper only the fixed-energy properties of random Hamiltonians came from an observation that such analysis can be made very elementary, even for multi-particle systems considered as difficult since quite a long time (cf. recent works \cite{CS08}, \cite{CS09a}, \cite{AW09}, \cite{CS09b}).


\section{The models and basic notations}\label{SectGeom}

A popular form of a single-particle random Hamiltonian is as follows:
$$
H(\om) = H^{(1)}(\om) = -\Delta + gV(x;\om),
$$
where $\Delta$ is the nearest-neighbor lattice Laplacian:
$$
(\Delta f)(x) = \sum_{y: \, \|y-x\|=1} f(y), \; x,y\in \DZ^d,
$$
and $V(x;\om)$ acts as a multiplication operator on $\DZ^d$. For the sake of simplicity, the random field $\{ V(x;\om), x\in\DZ^d\}$ will be assumed IID,  although a large class of correlated random fields can also be considered. In this paper, we consider only the case of "large disorder", i.e., we assume that $|g|$ is sufficiently large. An IID random field on $\DZ^d$ is completely determined by its marginal probability distribution at any site $x\in\DZ^d$, e.g., for $x=0$. We assume that the marginal distribution function
of potential $V$ (defined by $F_V(t) = \prob{ V(0;\om) \leq t}, \; t\in\DR$) is H\"{o}lder-continuous:
\be\label{eq:V.Hoelder}
\forall\, t\in\DR, \, \forall\,\eps>0 \;  F_V(t+\eps) - F_V(t) \leq Const \, \eps^b,
\ee
for some $b>0$.

For $N>1$ particles, positions of which will be denoted by $x_1, \ldots, x_N$, or, in vector notations,
$\ux = (x_1, \ldots, x_N)\in\DZ^{Nd}$, we introduce an interaction energy $U(\ux)$. Again, for the sake of simplicity
of presentation, we assume that
$$
U(\ux) = U(x_1, \ldots, x_N) = \sum_{1 \le j < k \le N} U_2(\|x_j - x_k\|),
$$
where $U_2(r)$, $r\geq 0$, is a bounded two-body interaction potential. The $N$-particle Hamiltonian considered below will have the following form:
\be\label{eq:def.H.N}
H^{(N)}(\om) = \sum_{j=1}^N \left(  -\Delta^{(j)} + gV(x_j;\om) \right) + U(\ux).
\ee

Given a lattice subset $\Lam\subset\DZ^d$,
we will work with subsets thereof called boxes. It is convenient to
allow boxes $\Lam_\ell(u)\subset\Lam$ of the following form:
$$
\Lam_\ell(u) = \myset{x:\, \| x-u\| \leq \ell -1},
$$
where $\|\cdot\|$ is the sup-norm: $\|x\| = \max_{1\leq j \leq d} |x_j|$.  Further,
we introduce a notion of internal and external "boundaries" relative to $\Lam$:
$$
\ba
\pt^- \Lam_\ell(u) = \myset{x\in\Lam:\, \| x-u\| \leq \ell-1}, \\
\pt^+ \Lam_\ell(u) = \myset{x\in\Lam:\, \dist(x,\Lam_\ell(u)) = 1}.
\ea
$$
We also define the boundary $\pt\Lam_\ell(u)\in\Lam$ by
$$
\pt\Lam_\ell(u) = \myset{ (x,x'):\, x\in\pt^-\Lam_\ell(u),
x'\in\pt^+\Lam_\ell(u), \, \|x-x'\|=1}.
$$

Observe now that the the second-order lattice Laplacian has the form
$$
\Delta = \sum_{(x, x'):\, \|x-x'\|=1} \Gamma_{x,x'},
\; \Gamma_{x,x'} = \ketbra{\delta_x}{\delta_{x'}},
$$
so that $(\Gamma_{x,x'}f)(y) = \delta_{x,y} f(x')$. Given a box $\Lam_\ell(u)\subset \Lam$, the Laplacian $\Delta_\Lam$ in $\Lam$ with Dirichlet boundary conditions on $\pt^+ \Lam$ reads as follows:
$$
\Delta_\Lam = \left( \Delta_{\Lam_\ell(u)} \oplus \Delta_{\Lam_L(u)\setminus\Lam_\ell(u)}\right)
+ \sum_{(x, x')\in\pt \Lam_\ell(u)} ( \Gamma_{x,x'} +  \Gamma_{x',x}).
$$
Fix a finite subset $\Lam$ and a box $\Lam_\ell(u)\subset \Lam$. Then for any point
$y\in(\Lam \setminus \Lam_\ell(u)$,  by the  second resolvent identity combined with the above decomposition, we obtain the so-called Geometric Resolvent Identity,
$$
G(u,y;E) = \sum_{(x, x')\in\pt \Lam_\ell(u)} \; G(u, x;) \, G(x',y;E).
$$
yielding the Geometric Resolvent Inequality (GRI):
$$
|G(u,y;E)|  \leq \Big(\,\mymax{ x \in \pt^- \Lam_\ell(u)} \, |G(u, x;E)| \cdot
 |\pt^+ \Lam_\ell(u)|   \,\,\Big)  \cdot
\mymax{x' \in \pt^+ \Lam_\ell(u)} |G(x',y;E)|.
$$

Throughout this paper, we use a standard notation $[[a,b]] := [a,b]\cap \DZ$.

\section{Green functions in a finite volume}\label{FVA}

\begin{definition}\label{DefS}
A box $\Lam_L(u)$ is called $(E,m)$-non-singular ($(E,m)$-NS) if
$$
\sum_{(y,y')\in\pt \Lam_L(u)} \; |G(u,y;E)| \leq e^{-\gam(m,L)L},
$$
where
\be\label{eq:def.gamma}
 \gam(m,L) = m\left(1 + L^{-1/4}\right).
\ee
Otherwise, it is called $(E,m)$-singular ($(E,m)$-S).
\end{definition}

\textbf{Remark.} The function $\gam(m,L)$ defined in Eqn \eqref{eq:def.gamma} will be often used below. It allows us to avoid a "massive rescaling of the mass", which would, otherwise, inevitably make notations and assertions more cumbersome. Obviously, $\gam(m,L)>m$, and for large values of $L$, $\gam(m,L)/m \approx 1$.

It is convenient to introduce the following property (or assertion), the validity of which depends upon parameters $u\in\DZ^d, L\in\DN^*, m>0, p>0$, as well as upon the probability distribution of the random potential $\{V(x;\om), x\in\DZ^d\}$:
\be\label{eq:SS.L.k}
\sssk{L,m} \quad \qquad \qquad \;
\prob{ \Lam_{L}(u) \text{ is } (E,m)\text{-S} } \leq L^{-p}.
\qquad \qquad
\qquad \qquad \quad
\qquad
\ee
In order to distinguish between single- and multi-particle models, we will often write \sssk{L,k,N}
where $N\geq 1$ is the number of particles.

\begin{definition}\label{DefCNR}
A lattice subset $\Lam\subset \DZ^d$ of diameter $L$ is called $E$-non-resonant ($E$-NR) if $\|G_\Lam(E)\|\leq e^{-L^{\beta} }$,
$\beta \in(0,1)$, and $E$-resonant ( $E$-R), otherwise. It is called $(E,\ell)$-completely non-resonant ($(E,\ell)$-CNR) if it is $E$-NR and does not contain any $E$-R cube $\Lam'$ of diameter $\geq 3\ell$.
\end{definition}

\begin{lemma}\label{LemWS1} If the marginal CDF $F(s) = F_V(s)$ is H\"{o}lder-continuous, then
$$
\exists\,L^*>0,  \beta' \in(0,1):\; \forall\, L\geq L^*\;
\prob{\Lam_{L}(u) \text{ is not } (E, L^{2/3})\text{-CNR} } \leq e^{-L^{\beta'}}.
$$
\end{lemma}


Consider a pair of boxes  $\Lam_\ell(u) \subset \Lam_L(x)$. If $\Lam_\ell(u)$ is $(E,m)$-NS, then GRI implies that function $f(x) = f_{y}(x) := G(x,y;E)$ satisfies
\be\label{eq:GF.subh.NS}
|f(u)|  \leq \qt  \cdot \mymax{v:\, \|v-u\|=\ell} |f(v)|,
\ee
with
\be\label{eq:qt}
\qt = \qt(d,\ell; E) = 2d\ell^{d-1} e^{-\gam(m,\ell)\ell}.
\ee

Now suppose that  $\Lam_\ell(u)$ is $(E,m)$-NS, but $\Lam_L(x)$ is $E$-CNR and, in addition,
for some $A>0$ and for any
$w$ with $\dist(w, \Lam_{A\ell}(u))=\ell$ the box $\Lam_\ell(w)$ is $(E,m)$-NS. Then, by GRI applied twice,
$$
\begin{array}{l}
|G(u,y)| \le 2d (6\ell)^{d-1} e^{L^\beta} \; \mymax{w: \|w-u\|=(A+1)\ell} |G(w,y;E)| \\ \\
\le 2d (6\ell)^{d-1} e^{L^\beta} \;2d (2\ell)^{d-1} e^{-\gam(m,\ell)\ell} \;
\; \mymax{v: \|v-u\| \in[A\ell, (A+2)\ell]} |G(v,y;E)|.
\end{array}
$$
Therefore, with
\be\label{eq:q.d.l.E}
 q(d,\ell,E) := 4d^2 (12\ell^2)^{d-1} e^{L^\beta} e^{-\gam(m,\ell)\ell} > \qt(d,\ell,E),
\ee
we obtain
\be\label{eq:GF.subh.S}
|G(u,y)| \le \qt(d,\ell,E) \; \mymax{v: \|v-u\| \in[A\ell, (A+2)\ell]} |G(v,y;E)|.
\ee
\smallskip

With these observations in mind,  we study in the next section decay properties of functions $f:\Lam_L(x)\to \DC$ obeying, for any point $u\in\Lam_L(x)$, one of the Eqns \eqref{eq:GF.subh.NS}, \eqref{eq:GF.subh.S}. Observe that we do not require that $q$ and $\qt$ be smaller than $1$, but, of course, the above bounds  are useful only for $q, \qt < 1$. Finally, note that, since $q<\qt$, it will be convenient to replace $\qt$ by $q$ in the bound (3.2). With this modification, we see that the only difference between bounds \eqref{eq:GF.subh.NS} and \eqref{eq:GF.subh.S} is in the distance $\|v-u\|$ figuring in these inequalities.

\section{Radial descent: A few simple lemmas }
\label{sec:rad.descent}

\begin{definition}\label{DefSubH}
Consider a cube $\Lam=\Lam_L(u)\subset\DZ^d$ decomposed into complementary subsets $\cS$, $\cR=\Lam\setminus\cS$,
and a  function $f:\,\Lam\to\DC$. Let $\ell\in(1,L-1)$ be an integer and $q>0$. Function $f$ will be called $(\ell,q,\cS)$-subharmonic  if for any $x\not\in\cS$ with $\dist(x, \pt \Lam)\ge \ell$, we have
\be\label{eq:def.subh.1}
|f(x)| \leq q \;\;\mymax{y:\, \|y-x\| \le \ell} |f(y)|
\ee
while for any $x\in\cS$ with $\dist(x, \pt \Lam)\ge \ell$
\be\label{eq:def.subh.2}
 |f(x)| \leq q \;\;\mymax{y:\, \|x - y\| \le r(x)+\ell} |f(y)|,
\ee
where
\be\label{eq:def.r.x}
r(x) = \min\{ r\in\DN:\,  \Lam_{r+\ell}(x) \setminus \Lam_{r-\ell}(x) \subset \cR \},
\ee
provided that the set of values $r$ in the RHS is non-empty. In all other cases, no specific upper bound on $|f(x)|$ is assumed.
\end{definition}

In other words, for a point $x$ in the "singular" subset $\cS\subset\Lam_L(u)$, $r(x)$ is the minimal radius of the "sphere" $C_r:=\{ \|y-x\| = r\}$ centered at $x$ and such that every ball $\Lam_\ell(y)$ with $y\in C_r$ is a subset of the "regular" subset $\cR\subset\Lam_L(u)$. Clearly, the difference between the upper bounds \eqref{eq:def.subh.1} and \eqref{eq:def.subh.2} resides in the shape of the "reference" set used for the calculation of the maximum in the RHS.

Note that, taking into account inequality \eqref{eq:GF.subh.S}, one could replace in the RHS of \eqref{eq:def.subh.2} the ball $\{y:\, \|y-x\| \le r(x) + \ell\}$ by a properly chosen annulus; however, this would not improve the final result on subharmonic functions, while making notations more cumbersome.

 We will use the notation $\cM(f, \Lam) := \max_{x\in \Lam} |f(x)|.$
Our goal is to obtain an upper bound on the value $f(x)$ exponential in $L/\ell$.

\begin{lemma}\label{lem:SubH.1}
Let $f$ be an $(\ell,q,\cS)$-subharmonic function on $\Lam = \Lam_L(x)$.
Suppose that $\cS$ can be covered by $K\ge 0$ cubes $\cS_1, \ldots, \cS_K$ with
$\sum_i \diam(\cS_i)\le \cW(\cS)$. Then
\be\label{eq:lem.SubH.1}
 |f(x)| \leq q^{[(L-\cW(\cS))/\ell]} \cM(f,\Lam).
\ee
\end{lemma}

The proof of Lemma \ref{lem:SubH.1} is given in Appendix C.

\subsection{From resolvents to subharmonic and monotone functions}

Lemma \ref{lem:SubH.1} will be applied in a situation where for  for any $E\in I$, a cube $\Lam_L(u)$ does not contain any collection of $K$ (or more) $(E,m)$-S cubes $\Lam_\ell(v^{(i)}$, $j=1, \ldots, K$ which are pairwise $a\ell$-distant, with $a\ge 0$:
$$
\forall\, i\ne j\;\; \dist(\Lam_\ell(v^{(i)}, \Lam_\ell(v^{(j)}) > 2a\ell
$$
In applications to the single-particle MSA with an IID potential $V$ it suffices to take $a=0$, while in the multi-particle MSA, or in the case where the potential $V$ has non-trivial but decaying correlations, one may need to use $a>0$ or even replace the RHS by $a\ell^{\delta}$ for some $1 <\delta < \alpha$.

\begin{lemma}\label{lem:G.subh.monot.1}
Consider a cube $\Lam_L(u)$, $L = [\ell^\alpha]$, and suppose that the random operator $H_{\Lam}(\om)$ fulfills the following conditions:
\begin{enumerate}[\rm (A)]
  \item $\Lam_L(u)$ contains no family of $K+1$ pairwise $a\ell$-distant $(E,m)$\emph{-S} cubes of radius $\ell$, for some $a>0$;
  \item $\Lam_L(u)$ is $E$\emph{-CNR};
\end{enumerate}
Then for any $w\in\pt \Lam$, the function
\be\label{eq:subH.def.f}
\begin{array}{lcll}
\funmapto{f }{ \Lam_{L}(u) }{ \DR }{ x }{ |G_{\Lam_L(u)}(x,w;E)| }
\end{array}
\ee
is $(\ell, q, \cS)$-subharmonic with
$
q \le c(d) (a\ell)^{d-1 }e^{-m\ell}
$
and the set $\cS$ contained in a union of $K'\le K$ cubes of radius $a\ell$ .
\end{lemma}
\proof

Fix some $E\in I$ and consider some maximal family of $(E,m)$-S cubes $\Lam_\ell(v^{(i)})$, $i=1, \ldots, K'\le K$; such a family may not be unique, but the maximal cardinality $K = K(I)$ is well-defined. By construction, any cube $\Lam_\ell(y)$ disjoint with the union of cubes $\Lam_{a\ell}(v^{(i)})$ is $(E,m)$-NS.

Pick a point $x\in\cS$ and consider the minimal cube $\Lam_{R(x)}(x)$ such that all cubes
$\Lam_{\ell}(y)$ with $y\in\Lam_{R(x)}(x)$  are $(E,m)$-NS. Assume  that the cube $\Lam_L(u)$ is $E$-CNR; then $\Lam_{R(x)}(x)$  is $E$-NR, and set $r(x) = R(x) + \ell$.

Applying the GRI twice and using the $E$-NR property of  $\Lam_{R(x)}(x)$, we obtain for any $w$ with $\| w - x\| > R(x)+\ell+1$:
$$
\ba
\diy |G_{\Lam_L(u)}(x, w;E)|  \\
\diy \le  d(2R(x)+1)^{d-1}) e^{(R(x))^{\beta}}   \max_{ \|y - x\|=R(x)}
\max_{ |v-y\|=\ell+1}
 \, |G_{\Lam_L(u)}(v, w;E)| \\
\diy \le  C(d) (a\ell)^{d-1}) e^{(a\ell)^{\beta}} e^{-\gam(m,\ell)\ell}
\max_{v:\, |v-x\| \le r(x)}
|G_{\Lam_L(u)}(v, x;E)|.
\ea
$$
Therefore, the function $f$ given by \eqref{eq:subH.def.f} is $(\ell, q, \cS)$-subharmonic with
\be\label{eq:subh.def.q.cS}
\ba
q = C(d) (a\ell)^{d-1}) e^{(a\ell)^{\beta}} e^{-\gam(m,\ell)\ell} \\
\cS = \cup_i \Lam_{a\ell}(v^{(i)}). \\
\ea
\ee
\qedhere

\section{Inductive bounds of Green functions}
\label{sec:Ind.GF.1part}

Here we summarize the results of \cite{S88} presented in a slightly modified form, making use of the notion of $(\ell,q)$-subharmonicity. Formally, we will need these results in Section \ref{SecMPMSA}, in the course of induction on the number of particles $N$.

\begin{lemma}\label{CorGRI}
Suppose that a box $\Lam_L(u)$  is $E$-NR and does not contain any $(E,m)$-S boxe of size $\ell$.
Then, with $q$ defined in Eqn \eqref{eq:subh.def.q.cS},
$$
\mymax{y\in\pt^- \Lam_L(u)} |G_{\Lam}(u,y;E)|
\leq q^{[L/\ell])} \, \|G_\Lam(E) \|.
$$
In particular, if $\ell^{-1}$ and $\ell/L$ are sufficiently small, so that
$$
\ell^{-1/4} - 6\ell L^{-1} \geq L^{-1/4},
$$
then
$$
 \sum_{y\in\pt \Lam_L(u)}|G_{\Lam}(u,y;E)|
\leq e^{ - \gam(m, L)L} \, \|G_\Lam(E) \|.
$$
\end{lemma}
\proof
Apply Lemma \ref{lem:G.subh.monot.1}. \qed

\begin{lemma}\label{lem:one.box}
Suppose that a box $\Lam_L(u)$  is $E$-CNR and does not contain any pair of
non-overlapping $(E,m)$-S boxes of size $\ell$.
Then
$$
\sum_{y\in \pt \Lam_L(u)} |G_{\Lam_L}(u,y)| \leq e^{-\gam(m,L)L}.
$$
\end{lemma}

\proof
If $\Lam_L(u)$ contains no $(E,m)$-S box $\Lam_\ell(x)$, the assertion follows from Lemma \ref{CorGRI}. Suppose there exists a box $\Lam_\ell(x) \subset \Lam_L(u)$ which is $(E,m)$-S. Since all boxes $\Lam_\ell(w)$ with $\|w-x\|>2\ell$ are disjoint with $\Lam_\ell(x)$, none of them is $(E,m)$-S, by hypotheses of the lemma. Therefore,  Lemma \ref{lem:SubH.1} applies, giving the required upper bound. \qed

\begin{lemma}\label{LemSMSA}
Suppose that the finite-volume approximations $H_{\Lam}(\om)$, $\Lam\subset \DZ^d$, $|\Lam|<\infty$,
of random LSO $H(\om)$ with IID random potential satisfy the following hypotheses:
\begin{itemize}
  \item \Wone $\;\prob{ \dist\left[E, \Sigma\left( H_{\Lam}(\om)\right) \right] \leq e^{-|\Lam|^{-\beta}}} \leq e^{-|\Lam|^{-\beta'}}$,
      for some $\beta, \beta'>0$;
  \item \s{\ell}
$  \prob{ \Lam_\ell(v) \text{ is } (E,m)\emph{-S} } \leq \ell^{-p}. $
\end{itemize}

\noindent
Set $L = [\ell^{3/2}]$. If $p>6d$, then $H_{\Lam_{L}}$ satisfies \ss{L,m,1}
(cf. \eqref{eq:SS.L.k}).
\end{lemma}

\proof
Consider a box $\Lam_{L}(u)$. By Lemma \ref{lem:one.box}, it must be $(E,m)$-NS, unless one of the following events occurs:
\begin{itemize}
\item $\Lam_{L}(u)$  is not $E$-CNR
\item $\Lam_{L}(u)$  contains at least two non-overlapping  $(E,m_k)$-S boxes
$\Lam_{\ell}(x), \Lam_{\ell}(y)$.
\end{itemize}
The probability of the former event is bounded by \Wone, while by virtue of \s{\ell},
the probability of the latter event is bounded by
$$
\qquad\qquad\qquad\qquad\qquad
\begin{array}{l}
\half  L^{2d} \, \ell^{-2p} \leq
\half  L^{-p\left(\frac{2}{3/2} - \frac{2d}{p} \right)}
\leq L^{-p}.
\end{array} \qquad\qquad\qquad\qquad{}_{_{\qed}}
$$

\begin{cor}\label{cor:DSk.step.1p}
\ss{L_k,m,1} implies \ss{L_{k+1}, m,1} (provided that Wegner-type bound \Wone$\,$ holds true).
\end{cor}

\begin{theorem}\label{ThmSPMSA}
Under the assumption \eqref{eq:V.Hoelder}, for any $m>0$, $p>0$
and sufficiently large $|g|\ge g^*(m,p)$, $\sssk{L_k,m,1}$ holds true for all $k\geq 0$.
\end{theorem}
\proof
It suffices to establish $\sssk{L_0,m,1}$. For any fixed $L_0$ and arbitrarily large $\eta>0$, we have
$$
\prob{\exists\, x\in\Lam_{L_0}(u):\, |gV(x;\om) - E|\le 2\eta }
\le \sum_{x\in\Lam_{L_0}(u)}\prob{ |gV(x;\om) - E|\le 2\eta }
\tto{|g|\to\infty}{}{} 0,
$$
provided that the marginal probability distribution of the potential is continuous. So, for any $p>0$ and  $g$ large enough, the above probability is bounded by $L_0^{-p}$. If $\eta>\|\Delta\|$,
then, by min-max principle, $\dist(E, Spec(H_{\Lam_{L_0}(u)}))\ge \eta>0$, and by Combes--Thomas estimate,
$$
|G_{\Lam_{L_0}(u)}(x,y)| \le e^{- M(\eta)\|x-y\|},
\; M(\eta) \tto{\eta\to+\infty}{}{} + \infty.
$$
This leads to the assertion $\sssk{L_0,m,1}$ and, by induction using Corollary
\ref{cor:DSk.step.1p}, to the assertion of the theorem.
\qedhere

Recall Kubo formula for the static conductivity $\sigma(E)$  at energy $E$:
$$
\sigma(E) = \lim_{\eps \downarrow 0} \eps^2 \sum_{x\in\DZ^d} \|x\|^2
\esm{ |G(0,x; E+ i\eps)|^2 }.
$$
By taking limit $k\to \infty$, i.e., $L_k\to\infty$, Theorem \ref{ThmSPMSA} implies immediately
\begin{theorem}\label{ThmSingleConduct}
Under the assumption \eqref{eq:V.Hoelder} for an IID random potential, the conductivity in the single-particle  Anderson model is zero.
\end{theorem}

\section{Simplified multi-particle MSA}
\label{SecMPMSA}

In this section, we study decay properties of Green functions for an $N$-particle Hamiltonian
$H^{(N)}(\om)$ defined in Eqn \eqref{eq:def.H.N}. We stress that estimates given below are \textit{far from optimal};
they only show that for any given number of particles $N\geq 1$ and any $m>0$, there exists a threshold $g_N = g_N(m)>0$ for the disorder parameter $g$ such that if $|g|\geq g_N$, then Green functions
in the $N$-particle model (with a short-range interaction) decay exponentially with rate $m$. t is also important to realize that using the MSA, in its traditional form,  for an $N$-particle system
with large $N$ inevitably requires using large values of $g_N$. Indeed, this phenomenon occurs even
in the single-particle MSA in high dimension $d\gg 1$: the respective threshold $g_1 = g_1(m,d)\to\infty$
as $d\to\infty$. A direct inspection of  the conventional
MSA show that, typically, $V(\cdot;\om$, $g_1(m,d) = O(d)$.

Given a number $m>0$ and an integer $L_0>0$, we will define  a decreasing sequence of decay exponents $m^{(n)}$, $n\geq 1$ as follows:
$$
m^{(1)} = m; \; m^{(N)} = m^{(N-1)} - L_0^{-1/8}, \; N>1.
$$
It is clear that in order to have $m^{(N)}>0$, we have to assume that $m = m^{(1)}$ is sufficiently large (depending on $N$). In turn, this requires the thresholds $g_1, \ldots, g_{N-1}$ to be large enough, as explained above.

Compared to the single-particle MSA scheme presented in previous sections, we have to replace
the property \ss{L,I} by a different one,  \ms{L,I,N} given below, and to include in
\ms{L,I,N} the requirement that the decay of Green functions
holds for all $n'<N$:

\ms{L_k,I,N}: The property \ss{L_k,I,n} holds for  $1\leq n\leq N-1$,
$$
 \prob{ \Lam^{(n)}_{L}(\uu) \text{ is } (E,m^{(n)})\text{-S} } \leq L^{-p(n,g)},
$$
where $p(n,g)\to \infty$ as $|g|\to \infty$, $n=1, \ldots, N-1$.

\textbf{Remark.} For any  $k=0$ and any $n>1$, the validity of the above statement is proved exactly in the same way as for $n=1$. Then for $k>0$ it is reproduced inductively. Denoting
$P(N-1,g) = \min_{1\le n \le N-1} p(n,g)$, we see that, under the hypothesis  \mssk{L_k,I,N}, we have $P(N-1,g)\to\infty$ as $|g|\to\infty$.

 For the sake of notational simplicity, we consider in Lemma \ref{lem:NITRONS} below only partitions  $[[1,N]] = \cJ \coprod \cJ^c$
of the interval $[[1,N]]$ into two (non-empty) consecutive sub-intervals,  $\cJ=[[1,n']]$, $\cJ^c=[[n'+1,N]]$. In other words, we consider a union of two subsystems with particles $1,\ldots,n'$
and $n'+1,\ldots,N$, respectively.

We introduce the "diagonal" subset of $\DZ^{Nd}$
$$
\DD := \myset{ x = (x_1, \ldots, x_1), \, x_1\in\DZ^d  }.
$$

\begin{lemma}\label{lem:NITRONS}
Let $[[1,N]] = \cJ \cup \cJ^c$, $\cJ =[[1,n']]$, $\cJ^c = [[n'+1,N]]$ and consider a cube
$
\Lam^{(N)}_L(\uu) = \Lam^{(n')}_L(u') \times \Lam^{(n'')}_L(u''),
$
with  $u'=(u_1, \ldots, u_{n'})$, $u'' = (u_{n'+1}, \ldots, u_N)$. Suppose that

\begin{enumerate}[\rm(1)]
  \item the interaction  between these two subsystems vanishes:
\be\label{eq:lem.NITOONS.hyp.1}
\forall\, j\in\cJ, \forall\, i\in\cJ^c \; \| u_j - u_{i} \| > 2(L - 1) + r_0,
\ee
so that
\be
\forall\, v_j \in \Lam^{(1)}_L(u_j), \forall\, v_i \in \Lam^{(1)}_L(u_i)\;\;
U_2(v_j, v_i) = 0.
\ee

  \item $\Lam^{(N)}_L(\uu)$ is $E$-NR;

  \item\ms{L_k,I,N} holds true;

  \item for any eigenvalue $\mu$ of $H_{\Lam^{(n'')}_L(u'')}$, the box
  $\Lam^{(n')}_L(u')$ is $(E-\mu, m^{(N-1)})$-NS;

  \item for any eigenvalue $\lam$ of $H_{\Lam^{(n')}_L(u')}$, the box
  $\Lam^{(n'')}_L(u'')$ is $(E-\lam, m^{(N-1}))$-NS.
\end{enumerate}

Then $\Lam^{(N)}_L(\uu)$ is $(E,m^{(N)})$-NS.
\end{lemma}

The proof of Lemma \ref{lem:NITRONS} is given in  Section \ref{AppA}; it is fairly straightforward.

\begin{lemma}
Under the assumption {\rm(1)} and {\rm(3)} of Lemma \ref{lem:NITRONS}, the assumptions
 {\rm(4)} and {\rm(5)} hold true with probability $\ge 1 - 2(2L+1)^{(N-1)d} L^{-p(N-1,g)}$.
\end{lemma}
\proof
The samples of the potential in the sets
$\Pi \Lam^{(n')}_L(u')$ and in $\Pi \Lam^{(n'')}_L(u'')$ (which are disjoint, owing to \eqref{eq:lem.NITOONS.hyp.1}), are independent. Therefore, using the conditioning on the potential in $\Pi \Lam^{(n')}_L(u')$ and the  one obtains
$$
\prob{\exists \, \lam\in \Sigma(H_{\Lam^{(n')}_L(u')}): \text{$\Lam^{(n'')}_L(u'')$
 is $(E-\lam,m^{(N-1)})$-S } }
 \le |\Lam^{(n')}_L(u') | \, L^{-p(n,g)},
$$
where $p(n'',g) \ge p(N-1,g)$. Similarly,
$$
\prob{\exists \, \mu\in \Sigma(H_{\Lam^{(n'')}_L(u'')}): \text{$\Lam^{(n')}_L(u')$
 is $(E-\lam,m^{(N-1)}))$-S } }
 \le |\Lam^{(n'')}_L(u'') | \, L^{-p(N-1,g)}.
$$
\qedhere

We will call $N$-particle boxes $\Lam_L^{(N)}(u)$ admitting a decomposition described in Lemma
\ref{lem:NITRONS} \textit{decomposable}. It is easy to see that an $N$-particle box $\Lam_L^{(N)}(u)$ is non-decomposable if and only if the union of cubes
$$
\bigcup_{j=1}^N \Lam_{L+r_0/2}^{(1)} (u_j) \subset \DR^{Nd}
$$
is connected (for this argument, we identify $\Lam_\ell^{(1)}(u_j)$ with a cube in $\DR^d$ of side $2(\ell-1)$ with center $u_j$, by a slight abuse of notations). In turn, such a union is connected, then
$$
\forall\, j =2, \ldots, N \;\; \|u_j - u_1\| \le 2(L-1) + r_0,
$$
so that
$$
\dist(u, \DD) \leq \|u - (u_1, \ldots, u_1)\|\le 2N(L-1)+Nr_0  =: r_{N,L}.
$$
Clearly, if $\dist(u, \DD) > r_{N,L}$, then $\Lam_L^{(N)}(u)$ is decomposable.

\begin{lemma}\label{LemTun}
Fix an integer $N>1$ and suppose that \mssk{L_k,I,n} holds for all $n < N$.
If an $N$-particle box $\Lam^{(N)}(u)$ is decomposable, then
$$
\prob{ \Lam^{(N)}_{L_k}(u) \text{ is } (E,m){\rm-S}  } \leq L_k^{-P(N-1,g)} + e^{-L_k^\beta}.
$$
Therefore, for any $P>0$ and for all $|g|$, $L_0$ large enough, we have
$$
\prob{ \Lam_{L_k}^{(N)}(u) \text{ is } (E,m){\rm-S}  } \leq L_k^{-P}.
$$
\end{lemma}

\begin{proof}
Since $\Lam^{(N)}_{L_k}(\uu)$ is decomposable, operator $H^{(N)}_{\Lam^{(N)}_{L_k}(\uu)}$
admits representation (7.1)
 for some $n', n''<N$. By Lemma \ref{lem:NITRONS}, either the box
$\Lam^{(N)}_{L_k}(\uu)$ is $E$-R, which occurs with probability $\leq e^{-L_k^\beta}$, or one of the operators $H^{(n')}_{\Lam^{(n')}_{L_k}(u')}$, $H^{(n'')}_{\Lam^{(n'')}_{L_k}(u'')}$ is $(E,m_k)$-S; the latter
event has probability $\leq L_k^{-P(N-1,g)}$, by \mssk{L_k,I,N-1}.
\qed
\end{proof}

Now we are prepared to give a \textbf{\textit{half page proof}} of the main result of the fixed-energy MSA. For notational simplicity, below we write $m$ instead of $m^{(N)}$.

\begin{lemma}\label{MPMSAstep} For $|g|$ large enough,
\mssk{L_k,I,N} implies \mssk{L_{k+1}, I,N}.
\end{lemma}

\proof
By Wegner bound, $\prob{\Lam^{(N)}_{L_{k+1}} \text{ is not $E$-CNR} } \leq e^{-L_{k+1}^{\beta}}$. Further, set
$$
\uS = \myset{ \exists \, \Lam_{L_k}(x)\subset \Lam_{L_{k+1}}(u) \text{ which is $(E,m)$-S and }
\dist(\Lam_{L_k}(x), \DD) \le r_{N,L} }.
$$
By Lemma \ref{LemTun}, and for $|g|$ large enough,
$
\begin{array}{c}
\prob{ \uS }  \le \half L_{k+1}^{-p(N,g)}.
\end{array}
$
Next, set
$$
\begin{array}{l}
\uB = \Big\{ \exists \, \Lam_{L_k}(x), \Lam_{L_k}(y)\subset \Lam_{L_{k+1}}(u) \text{ which are $(E,m)$-S and such that } \\
\|x-y\| > 9 (L_k + r_{N,L_k}), \;\dist(\Lam_{L_k}(x), \DD) \le r_{N,L},
\; \dist(\Lam_{L_k}(y), \DD) \le r_{N,L} \Big\}.
\end{array}
$$
Fix centers $x, y$. By Lemma \ref{GeomDiag}, potential samples in boxes
$\Lam_{L_k}(x)$, $\Lam_{L_k}(y)$ close to the "diagonal" $\DD$ are
\textbf{\textit{independent}}, so that
$$
\prob{ \uB } \leq \half L_{k+1}^{2d} \left(\mymax{v\in\Lam_{L_{k+1}}(u)}
\prob{ \Lam_{L_k}(v) \text{ is $(E,m)$-S}}\right)^2  < \half L_{k+1}^{-p(N,g)}.
$$
If neither of events $\uS, \uB$ occurs, then, by Lemma \ref{LemSMSA},
$\Lam_{L_{k+1}}(\uu)$ is $(E,m_{k+1})$-NS. Therefore,
with $p(N,g)\to\infty \text{ as } |g|\to\infty$, we have
$$
 \prob{ \Lam^{(N)}_{L_{k+1}}(\uu) \text{ is } (E,m))\text{-S} }  \le
\half L_{k+1}^{-p(N,g)} + \half L_{k+1}^{-p(N,g)} = L_{k+1}^{-p(N,g)}. \;\;\;\;\qed
$$

\begin{theorem}\label{ThmMultiConduct}
Under the assumption \eqref{eq:V.Hoelder} for an IID potential, the conductivity in the $N$-particle Anderson tight binding model is zero.
\end{theorem}

\section{Appendix A}
\label{AppA}

\textit{Proof of Lemma \ref{lem:NITRONS}}

Let $\{ \ffi_a, \lam_a\}$ be  normalized eigenfunctions (EFs) and the respective eigenvalues (EVs) of $H^{(n')}_{\Lam^{(n')}(u')}$, and $\{ \psi_b, \mu_b\}$ be  normalized EFs and the  EVs of $H^{(n'')}_{\Lam^{(n'')}(u'')}$.  Due to absence of interaction between configurations $u'$, $u''$, we have
\be
H^{(N)}_{\Lam^{(N)}(\uu)} =  H^{(n')}_{\Lam^{(n')}(u')} \otimes \one + \one \otimes H^{(n'')}_{\Lam^{(n'')}(u'')}.
\ee
Therefore, the EFs $\Psi_{a,b}$  of operator $H^{(N)}_{\Lam^{(N)}(\uu)}$ have the form
$\Psi_{a,b} = \ffi_a \otimes \psi_b$, and the respective EVs are given by
$E_{a,b} = \lam_a + \mu_b$, so that $ E-E_{a,b} = \left(E-\lam_a \right) - \mu_b $.

Let $\uv\in\pt \BLam^{(N)}_L(\uu)$. Then either $\|u' - v'\|=L$, or
$\|u'' - v'' \|=L$.
Without loss of generality,
assume that $\|u' - v'\|=L$. Since  $\|\ffi_a \|=1$
and $n'<N$, we have
$$
\bal
 \left| \uG(\uu, \uv;E) \right| &\le
 \sum_{a} \left|\ffi_a(u') \ffi_a(v')\right| \,\left| \sum_{b} \, \frac{ \psi_b(u'') \psi_b(v'') }{ (E - \lam_a) - \mu_b } \right| \\
 &\le \left| \Lam^{(n')}_{L}(u') \right| \, \mymax{\lam_a}\, | G_{\Lam^{(n'')}_{L}(u'')}(u'', v''; E - \lam_a)|
 \le L^{N-1} \, e^{-L^\beta}  e^{-mL}. \qquad\qed
 \eal
$$

\section{Appendix B}
\label{AppB}
\def\ut{{\widetilde{u}}}
\def\vt{{\widetilde{v}}}

\begin{lemma}\label{GeomDiag}
Consider two points $x,y\in\DZ^{Nd}$ and an integer $ L >0$. Suppose that
$\| x - y \| > AL$ with $A = 4B+6$, $B>0$, and
\be\label{eq:App.B.1}
\dist(\Lam^{(N)}_L(x), \DD_0) < BL,  \quad
\dist(\Lam^{(N)}_L(y), \DD_0) < BL.
\ee
Then for all $j=1, \ldots, N$, we have
$
\Pi_j\Lam^{(N)}_{L}(x) \cap \Pi_j\Lam^{(N)}_{L}(y) = \emptyset.
$
\end{lemma}
\begin{figure}[h]
\center
    \includegraphics[height=0.50\textheight]{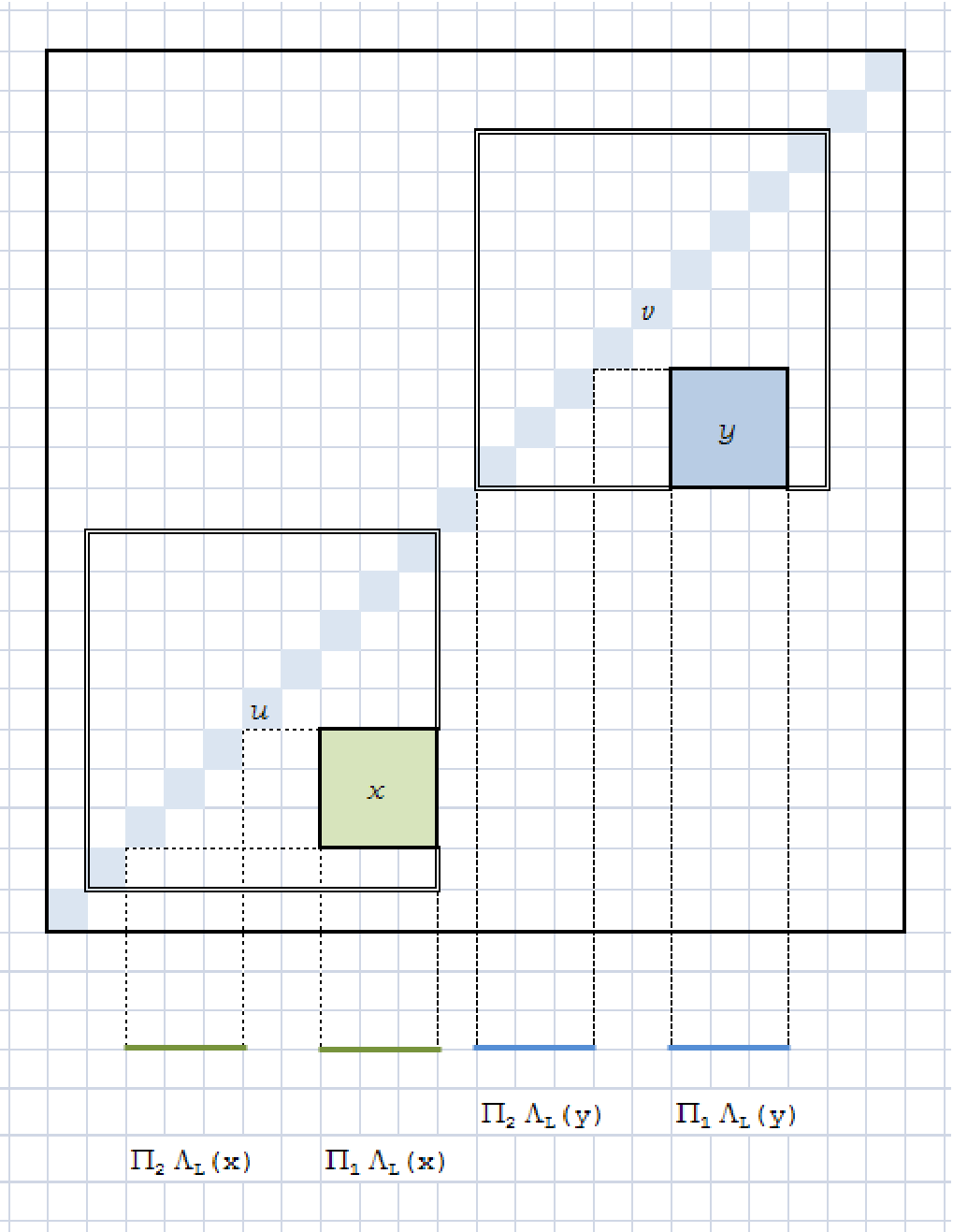}
    \caption{Cubes $\Lam^{(2)}_L(x)$ and $\Lam^{(2)}_L(y)$ with non-overlapping projections}\label{fig:fig04}
\end{figure}

\proof (See Fig. 4 illustrating the case $N=2$, $d=1$.)
Eqn~\eqref{eq:App.B.1} implies that there exist  some points $u = (u_1, \ldots, u_1)\in\DD$,
$v = (v_1,\ldots, v_1)\in\DD$
such that $\|x - u\|\le L + BL$, $\|y - v\|\le L + BL$. Then we have, by triangle inequality,
$$
\begin{array}{l}
\|x - y\|  \le \|x - u\|+  \|u - v\| + \|v - y\|
 \le (2B+2)L + \|u - v\|,
\end{array}
$$
so that
\be\label{eq:dist}
\|u - v\| \ge \|x - y\| - (2B+2)L > 2(B + 2)L,
\ee
and also
$$
\Lam_L(x)\subset \Lam_{(B+2)L}(u),
\; \Lam_L(y)\subset \Lam_{(B+2)L}(v).
$$
Similar inclusions hold true for the projections:
$$
\Pi_j\Lam^{(N)}_{L}(x)    \subset \Pi_j \Lam_{(B+2)L}(u),
\; \Pi_j\Lam^{(N)}_{L}(y) \subset \Pi_j  \Lam_{(B+2)L}(v).
$$
However, Eqn \eqref{eq:dist} implies that
cubes $\Lam_{(B+2)L}(u)$ and $\Lam_{(B+2)L}(v)$ are disjoint, and so are, therefore, the projections $\Pi_j\Lam^{(N)}_{L}(x)$ and $\Pi_j\Lam^{(N)}_{L}(y)$.
\qed

\section{Appendix C. Decay of subharmonic functions }
\label{sec:appendix.C.SubH}

\subsection{From subharmonic to monotone functions}

To emphasize the essentially one-dimensional nature of the argument used in the proof of Lemma \ref{lem:SubH.1} and to make notations and calculations less cumbersome,  we introduce the `radial', one-dimensional counterpart of the notion of an $(\ell,q,\cS)$-subharmonic function.

\begin{definition}
\label{def:monot.rad}
Consider an interval $I = [[0,L]]$ decomposed into complementary subsets $\cS$,
$\cR = I \setminus \cS$, and let a number $q>0$ and an integer $\ell\in[1, L/2)$ be given.
A non-negative function $F:I\to\DR_+$   will be called $(\ell,q,\cS)$-monotone,
if for any $r\in \cR\cap [[0,L-\ell]]$ one has
$$
F(r) \le q \cM(F, [[0, r+\ell]]),
$$
while for any $r\in \cS\cap [[0,L-\ell]]$
$$
F(r) \le q \cM(F, [[0, R(r)+\ell]]),
$$
where $R(r)$ is defined by
\be
R(r) = \min\{r'\in[[r+1,L]]:\, [[r'-\ell, r'+\ell]]\subset \cR \},
\ee
in the case where the set of values  $r'$ figuring in the RHS is non-empty. In all other cases, no specific upper bound on $F(r)$ is assumed.
\end{definition}


\begin{lemma}\label{lem:monot.2}
Let $F:[[0,L]]\to\DR_+$ be an $(\ell,q,\cS)$-monotone function. Suppose that $\cS$ consists of
$K\ge 0$ disjoint intervals of total length $W = |\cS|$. Then
\begin{enumerate}[\rm(A)]
  \item
\be\label{eq:lem.monot.claim.A}
F(0) \le q^{[(L- W)/\ell] } \cM(F, [[0,L]])
\ee
(with $[\,\cdot\,]$ in the exponent standing for the integer part);
  \item more generally, for any $0 \le r < L- W-\ell$,
\be\label{eq:lem.monot.claim.B}
\cM(F, [[0, r]])\le q^{[(L- r - W)/\ell]} \cM(F, [[0,L]]).
\ee
\end{enumerate}
\end{lemma}

\proof

First of all, observe that one can replace the interval $I=[[0,L]]$ by
$$
I' = [[0, \ell+\max\{ r\in I: \, [[r-\ell, r+\ell]]\subset \cR\}  ]],
$$
setting also $\cS' = \cS\cap I'$, $\cR' = \cR\cap I'$, and restricting $F$ on $I'$. Indeed, there is no assumption on the function $F$ for the points in $I\setminus I'$. Clearly, this gives rise in $K'\le K$ and $|\cS'|\le |\cS|$. We will assume that this reduction has already been performed.

Next, without loss of generality, assume that $\cS$ is a union of disjoint intervals $\cS_j$,
$j=1, \ldots, K$, and consider any of these intervals,  $\cS_j = [[a_j,b_j]]$. By definition, for any point
$s\in\cS_j$ we have
$$
F(s) \le q  \cM(F, [[0, R(s)+\ell]]),
$$
and $R(s)$ is constant on the given "singular" interval $\cS_j=[[a_j,b_j]]$; more precisely,
$$
R: [[a_j-1, b_j]] \to \{ R(a_j - 1) \}.
$$
It is convenient to set $R(r) = r$ for all $r\in\cR$, so that the new function $R$, defined now on the entire interval $I$, is monotone non-decreasing, and strictly increasing at each point $r\in\cR$. Introduce a mapping $\csX:I\to\DN$ by
$$
\csX(r) =  \card\{ [0,r]\cap \cR \}
$$
and a new interval $\tI = \csX(I)$; we will consider $\csX$ as a mapping $\csX: I\to\tI$. Note that
for each singular interval $\cS_j = [[a_j,b_j]]$ one has
$$
\csX: [[a_j-1, b_j]] \to \{ \csX(a_j - 1) \}.
$$
On the other hand, if $r'\in\cR$ and $r := \csX(r')$, then $\csX^{-1}(r) = \{r'\}$. As a result,
for any $r'\in I$, $\csX(r'+1) - \csX(r') \le 1$ and, more generally, for any $s' \le r'$,
\be\label{eq:cX.bdd.1}
\csX(r') - \csX(s') \le r' - s'.
\ee

Define a function $\tF:\tI\to\DR_+$ by
$$
\tF(r) = \max_{r'\in \csX^{-1}(r)} F(r').
$$
Let us show that $\tF$ is $(\ell,q)$-monotone on $\tI$ with the same values of $q$ and $\ell$ as for the original function $F$. To this end, consider $r_0\in\tI$ and any point
$r'_0\in\csX^{-1}(r_0)$.
The $(\ell,q)$-monotonicity of function $F$ implies that
$$
F(r'_0) \le q \cM(F, [[0, R(r'_0)+\ell]]).
$$
Set
$$
r'' = \max_{r\le r_0 + \ell } \; \max \{ r' \in \csX^{-1}(r) \} \in I.
$$
Then $r'' \ge r'_0 + \ell$, since, by Eqn \eqref{eq:cX.bdd.1},
$$
r'' - r'_0 \ge \csX(r'') - \csX(r'_0) = (r_0+\ell) - r_0 = \ell.
$$
Combining this inequality with $R(r'_0) \ge r'_0$, one can write
$$
\bal
F(r'_0) & \le q \cM(F, [[0, R(r'_0)+\ell]]) \\
        & \le q \cM(F, [[0, r'_0+\ell]]) \\
        & \le q \cM(F, [[0, r'']]) =  q \cM(\tF, [[0, r_0+\ell]]).
\eal
$$
Since this bound is valid for every point $r'_0\in\csX^{-1}(r_0)$, we conclude that
$$
\tF(r_0)\le q \cM(\tF, [[0, r_0+\ell]]).
$$
For the $(\ell,q)$-monotone function $\tF:\tI\to\DR_+$, one easily obtain by induction
$$
\cM(\tF, [[0, \csX(L) - n\ell]]) \le q^n \cM(\tF, [[0, \csX(L)]]) = q^n \cM(F, [[0, L]]),
$$
as long as $\csX(L) - n\ell \ge 0$. Clearly, the maximal possible value of $n$ is given by
$$
n = [\csX(L)/\ell] = \left[ \frac{L - W} {\ell} \right].
$$
This completes the proof of assertion (A).

Assertion (B) follows from (A) applied to the interval $[[r,L]]$.
\qedhere

\subsection{Proof of Lemma \ref{lem:SubH.1}}

$\,$

Introduce the function
$$
\begin{array}{llll}
\funmapto{F}{ \Lam }{ \DR_+}{ x }{ \cM( |f|, \rS_{\|x\|})}.
\ea
$$
It is not difficult to see that it is $(\ell,q,\cSt)$-monotone, with
$$
\cSt := \{r:\, \rS_r \cap \cS \ne \varnothing\},
$$
so that the claim follows from Lemma \ref{lem:monot.2}.
\qed

\section*{Acknowledgments}

I would like to thank the Isaac Newton Institute (INI) and the organizers of the INI programs IGCS (2003), AGA (2007) and MPA (2008) for their kind invitations and warm hospitality during my stay at the INI.
It is my pleasure to thank also many participants to these programs, especially MPA, for numerous fruitful discussions. In fact, the main idea of this short note is a direct result of my discussions with Boris Shapiro. I regret that Boris did not find possible to become a co-author of the present paper; it does not make, however, his contribution less important. Special thanks to Tom Spencer for initiating my collaboration with Boris Shapiro and many stimulating discussions, and also to Y. Fyodorov, A. Mirlin, D. Shepelyanski and S. Fishman (in chronological order of discussions) for sharing with me their physical insights.

I also thank Yuri Suhov for  discussions of a preliminary version of this short note.


\end{document}